\documentclass[oldversion]{aa}  
\usepackage{graphicx}
\usepackage{txfonts}
\def\ef{EF\,Eri}
\def\crat{s$^{-1}$}
\def\porb{P_{\rm orb}}
\def\msun{M$_\odot$}

\def\lum{erg\,s$^{-1}$}
\def\ftot{erg\,cm$^{-2}$\,s$^{-1}$}
\def\flam{erg\,cm$^{-2}$\,s$^{-1}$\,\AA$^{-1}$}

\def\teff{T_{\rm eff}}

\begin{document}
   \title{XMM-Newton observations of EF Eridani: the textbook example of
   low-accretion rate polars\thanks{Based on observations obtained with
   XMM-Newton, an ESA science mission with instruments and contributions
   directly funded by ESA Member States and NASA}}

   \author{A.D. Schwope\inst{1}
     \and A. Staude\inst{1}
     \and D. Koester\inst{2}
     \and J. Vogel\inst{1}
   }

   \institute{Astrophysikalisches Institut Potsdam,
                An der Sternwarte 16, D-14482 Potsdam\\
              \email{aschwope@aip.de}
              \and
              Institut f\"ur Theoretische Physik und Astrophysik, 
              Universit\"at Kiel, 24098 Kiel, Germany}

   \date{Received ; accepted }

\abstract{Archival X-ray observations of EF Eridani
  obtained in a low state revealed
  distinct X-ray detections at a luminosity $L_X \simeq 2\times 
  10^{29}$\,erg\crat, three orders of magnitude below its high state
  value. The plasma temperature was found to be as low as $kT \loa 2$\,keV,
  a factor 10 below the high state. The X-ray/UV/IR spectral energy
  distribution suggests faint residual accretion rather than coronal emission
  as being responsible for the low-state X-ray emission. \ef\ thus showed a
  clear transition from being shock-dominated in the high state to be
  cyclotron-dominated in the low state. 
  From the optical/UV spectral energy distribution we 
  re-determine the photospheric temperature of the white dwarf to 
  $\sim$10000\,K. Contrary to earlier claims,
  WD model atmospheres produce sufficient UV flux to reproduce the
  published {\it GALEX} flux and orbital modulation.
}
\maketitle

\section{Introduction}
Polars are magnetic cataclysmic binaries consisting of a late-type 
main-sequence star and a strongly magnetic white dwarf locked in synchronous 
rotation. \ef\ was one of the 11 polars known in the pre-ROSAT era, it was the
second brightest at optical and at X-ray wavelengths
after the prototypical system AM Herculis. It was studied with
all major X-ray observatories (EINSTEIN, EXOSAT, GINGA, ROSAT) in the past
and was always found in a high accretion state.

EINSTEIN observations revealed the presence of uncorrelated soft and hard
X-ray emission and were used to observationally establish the standard picture
of magnetic accretion onto white dwarfs in the high $\dot{m}$-regime dominated
by a shock-heated accretion column and cooling by free-free radiation
(Beuermann, Stella \& Patterson~1987, henceforth BSP87).
The absence of a pronounced soft X-ray excess made BSP87 to coin
\ef\ the textbook example of AM Herculis-type systems.  
The shape of the X-ray light curves, in particular the presence of a soft
X-ray absorption dip, was used to uncover the accretion geometry. \ef\ had a
main accretion pole which was continuously in view, the observer has a
moderate inclination with respect to the orbital plane, so that the line of
sight crosses the accretion stream on its way through the magnetosphere. 
This special geometry allowed detailed stream-density diagnostics with GINGA
and EXOSAT (Watson et al.~1989). 

The accretion geometry was intensively studied using photo- and
spectro-polarimetric data (e.g.~Bailey et al.~1982, Cropper 1985,
Piirola et al.~1987, Meggitt \& Wickramasinghe 1989, Beuermann et al.~2007).
The latter three papers agree that the white dwarf's magnetic
field is probably more complex than that of a centered dipole.
The zero point of Bailey's ephemeris (Bailey et al.~1982)
centered on the IR (X-ray) absorption dip is widely used in the
literature. Piirola et al.~(1987) determined an updated orbital
period based on a linear regression of the arrival times of linear
polarisation pulses. 
Beuermann et al.~(2007) derived a slightly revised ephemeris by 
including the ROSAT PSPC X-ray dip timings from July 1990.
Phases in this paper refer to Bailey's phase zero and 
Piirola's period. 

\ef\ turned into a deep low state at $V\simeq 18$ in 1997 (Wheatley \& Ramsay
1998) and remains therein since then. 
A re-brightening was reported in VSNET
 on March 5, 2006 (ERI EF 20060305.724 at 14.2 unfiltered CCD based on the
 Henden-Sumner sequence), but the system returned to the low state shortly
 thereafter. 
While in the high state the stellar
photospheres are outshone by accretion radiation, the low state offers the
opportunity to investigate the stars, at least in principle. 
Since \ef\ is the polar with the shortest orbital period,
$P_{\rm orb} =81$\,min, just a few minutes above the CV minimum period, low
state observations are of utmost importance to test current scenarios of CV
evolution and to search for the cool secondary. Indeed, following the more 
indirect conclusion by Beuermann et al.~(2000) of a substellar secondary in
\ef\ from the non-detection of any spectral signature of the companion in
optical spectra, Howell \& Ciardi (2001) claimed the detection of the
secondary in near-infrared spectra. A more 
likely explanation of the observed infrared humps was given in terms of
cyclotron radiation (Beuermann et al.~2000, Harrison et al.~2004). 

Beuermann et al.~(2000) also estimated the photospheric temperature of the
white dwarf from their low-resolution optical spectra, $T_{\rm WD}
=9500\pm500$\,K, one of the coldest WDs among all CVs. This allowed to draw
some conclusions on the likely evolutionary state of the object. 
Recently, Szkody et al.~(2006) report on phase-resolved {\it GALEX}
observations with the puzzling result of a distinct source of ultraviolet 
flux much larger than the underlying 9500\,K white dwarf. 

Here we report on archival XMM-Newton observations of \ef\ obtained in a low
accretion state. We search for remaining X-ray emission in the low state
either originating from the white dwarf or the 
secondary and analyse the data from the optical monitor taken through two
different filters.

\section{Low-state observations with XMM-Newton}
The XMM-Newton Science Archive (XSA) contains three observations of the X-ray
sky in the direction of \ef. They are listed with their nominal exposure times
in Tab.~\ref{t:log}. 

\begin{table}[t]
\caption{XMM-Newton observations of \ef. The first column lists the unique
  observation ID and the revolution number of the spacecraft, the last column
  lists the nominal and effective 
  exposures times of the individual observations, the latter quantity after
  screening for high background and other instrumental defects.}
\label{t:log}
\begin{tabular}{lccc}
\hline\hline
OBSID/rev & Date & Instr. & Exp Nom/Eff\\
& & & (s)\\
\hline
0111320201/496 & 2002-08-24 & EPIC-PN & 6132/1859\\
& & OM V & 6000 \\
0111320401/571 & 2003-01-20 & EPIC-MOS & 5160/5088\\
0111320501/583 & 2003-02-14 & EPIC-PN & 5047/4559\\
& & EPIC-MOS & 6660/6575 \\
& & OM V & 3400 \\
& & OM UVW1 & 2600\\
\hline
\end{tabular}
\end{table}

We refer to individual observations by an 'E' followed by the last three
digits of the OBSID, i.e.~E201 for the observation on 28$^{\rm th}$ of April,
2002. All X-ray observations were obtained in full frame mode. The RGS did not
reveal useful data due to low count rate and will not considered further.
Data processing was performed with the latest version of the XMM-Newton SAS
(version 7.0), a spectral analysis of the X-ray data was performed with
XSPEC. Despite the relatively short exposure times, almost full phase coverage
of the $\porb = 81$\,min binary was achieved at two occasions (E401 and E501).
The accumulated phase uncertainty of the period derived by Piirola et
al.~(1987) at the epoch of the last XMM-Newton X-ray observation, i.e. after
$\sim$155000 binary cycles, is 0.014 phase units only and thus negligible. 

EPIC-MOS and EPIC-PN show a faint apparent companion to \ef\ 
at $\alpha(2000) = 03^h 14^m 14\fs0$ and $\delta(2000) = -22\degr 36' 04''$. 
The source has no counterpart on DSS2 images. 
It contributes at a level of $F_X \simeq  1.5 \times 10^{-14}$\,\ftot\ in the
0.1 -- 10 keV band. This source could not be resolved in all previous X-ray
observations with other satellites. Its faint flux was just a small 
contamination of all previous X-ray observations and thus irrelevant. 
It represents, however, a $\sim$20\% contamination of the flux of \ef\ during
XMM-Newton observations. We thus chose source and background regions for the
extraction of light curves and spectra avoiding the region around this source.

\subsection{X-ray spectra and light curves}
The net exposure time of observation E201 was just 1859\,s. The source was
detected with EPIC-PN at a mean count rate of 0.022\,\crat. The spectrum
contains no
photons above 5 keV, it could be successfully fitted (reduced $\chi^2 = 0.93$
for 9 degrees of freedom) with a cooling plasma model (MEKAL in XSPEC terms)
with a temperature of $kT = 2.8\pm1.7$\,keV only very little affected by
interstellar absorption. 
In general, all X-ray spectral fits based on XMM-Newton
observations are compatible with zero interstellar absorption, in accord with
the low column density inferred from ROSAT and EXOSAT, 
$N_{\rm H} = 10^{19}$\,cm$^{^2}$ (Beuermann et al.~1991; Watson, King \&
Williams 1987), and from EINSTEIN, $N_{\rm H} < 1 \times 10^{20}$\,cm$^{^2}$
(BSP87). 
The integrated flux in this component was 
$F_X = 7 \times 10^{-14}$\,\ftot\ (0.1 -- 10 keV).

Observation E401 was performed with EPIC-MOS only and resulted in the
detection of 26/20 photons in 5099/5088\,s with MOS1/2, respectively.
The spectrum was found to be very soft again, an unconstrained fit yielded 
$kT = 0.5 \pm 1$\,keV, but the spectral parameters remained highly uncertain
due to the small number of photons. A fit using the same parameters as for
E501, see below, yielded a flux of $F_X \simeq 4 \times 10^{-14}$\,\ftot\ (0.1
-- 10 keV), slightly indicative of a lower X-ray flux at that epoch. 

\begin{figure}[t]
\resizebox{\hsize}{!}{\includegraphics[angle=-90,clip=]{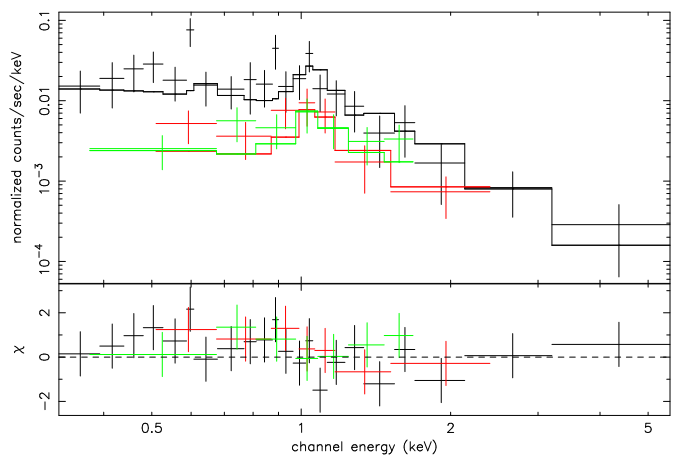}}
\caption{Mean X-ray spectrum of \ef\ (observation E501) and best-fit thermal
  plasma model. }
\label{f:spec501}
\end{figure}

Observation E501 was performed with all three X-ray cameras onboard 
and revealed 140, 43, and 40 source photons with EPIC-PN, MOS1, and MOS2,
respectively. The mean spectrum, which is a good approximation to the orbital
mean spectrum also, is shown in Fig.~\ref{f:spec501}. Again, it is a soft
spectrum which could be fitted with just one emission component (reduced
$\chi^2 = 0.81$ for 33 d.o.f.). The best-fit
plasma temperature of the MEKAL model is $kT = 1.7 \pm 0.2$\,keV, and the
integrated flux $F_X = 6 \times 10^{-14}$\,\ftot\ (0.1 -- 10 keV).
The $O-C$ residuals of such a fit (see Fig.~\ref{f:spec501}) 
give the slight impression of a systematic
trend with an excess of photons between 0.5--1.0\,keV. However, the parameters
of any additional spectral component cannot be constrained significantly and
we thus stick to a one component X-ray spectrum. 

We found no significant X-ray variability of the total X-ray flux between the
three XMM-Newton observations.
X-ray variability in E501, the longest of the three observations, was 
almost insignificant. A binned light curve with
bin size 243 s (20 phase bins per orbital cycle) shows one bin with no source
photon. It occurs at phase 0.7, i.e.~it cannot be associated with the high
state absorption dip (if the accuracy of Piirola's period is as high as the
formal uncertainties given in their suggest). Given the small number of 
X-ray photons a secure claim on the existence of a dip cannot be made.

\begin{figure}[ht]
\resizebox{\hsize}{!}{\includegraphics[angle=-90,clip=]{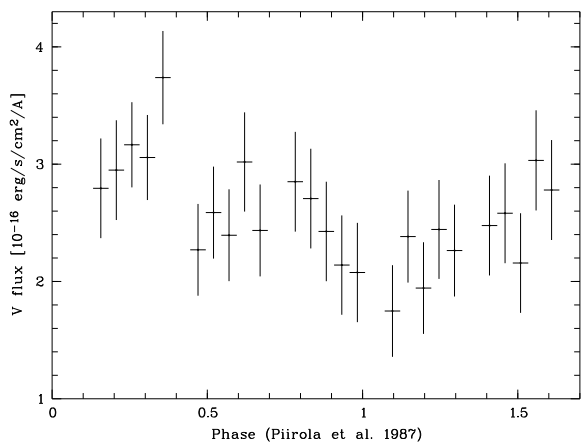}}
\caption{OM light curve through $V$-filter from observation E201. The bin
  size is 243\,s corresponding to 0.05 phase units.}
\label{f:omlc}
\end{figure}

\subsection{Optical/UV observations with the OM}
The optical monitor OM was used in observations E201 and E501 with the $V$
and $UVW1$ filters, respectively (see Tab.~\ref{t:log} for details). In E201
full phase coverage was achieved, in E501 only parts of the orbital cycle were
covered with the two filters. During E201, the mean countrate in the
$V$-filter (phase 0.16 -- 1.61) 
was 0.88(8)\,\crat\ corresponding to a mean flux of
$F_V = 2.6(2) \times 10^{-16}$\,\flam. During E501 the mean countrate through
the $V$-filter (phase interval 0.90 -- 1.55) was 0.87(5)\,\crat\
corresponding to $F_V = 2.2(1) \times 10^{-16}$\,\flam), 
and through the $UVW1$-filter (phase interval 0.67 -- 1.17) 
it was 1.32(4)\,\crat\ corresponding to $F_{UVW1} = 6.5(2) \times
10^{-16}$\,\flam. 

Some modulation of the optical flux at a level of about 50\% was discovered in
E201 (Fig.~\ref{f:omlc}) with a minimum around phase zero. The low-state light
curves by Skody et al.~(2006) show a minimum at phase 0.4. 
The phase difference between the two epochs by using either Bailey's period
used by Szkody et al.~(2006) and Piirola's period used here is negligible,
0.01 phase units. Hence, this phase shift, if a real feature of the light
curve, cannot be explained by different phase conventions. 
However, both data sets were 
obtained with small telescopes and are rather noisy. Given the rather large  
error bars on individual light curve bins we are not further discussing 
possible differences between the light curves.
We note, however, that the accumulated phase difference from Bailey's zero
point in 1979 to the epochs of the GALEX (2004) or XMM-Newton (2003)
observations is 0.18 phase units with either Bailey's or Piirola's period and
thus not negligible.  
Formally, the period derived by Piirola et al.~should be preferred due to the
claimed higher accuracy, but an independent re-determination of the linear
polarization ephemeris is highly desirable, should \ef\ ever return to a high
accretion state. 

\begin{figure}[t]
\resizebox{\hsize}{!}{\includegraphics[clip=]{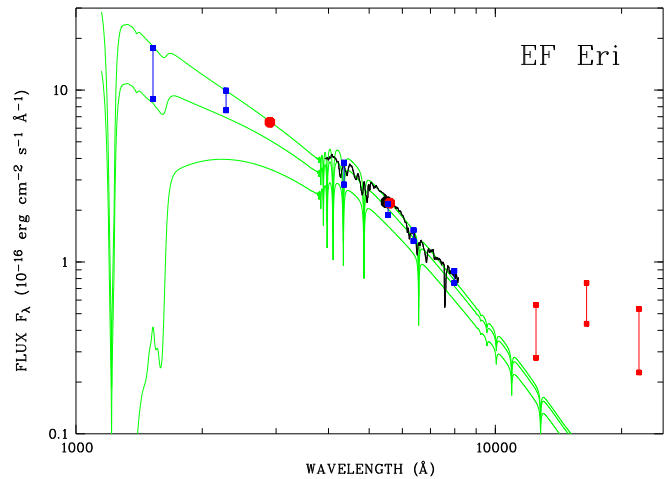}}
\caption{Ultraviolet to infrared spectral energy distribution of \ef\ in  the
  low state. Shown are an optical low-state spectrum and infrared JHK
  photometry adapted from Harrison et al.~(2004), GALEX-UV and optical BVRI 
  photometry from Szkody et al.~(2006, blue dots),
  OM V-band and UVW1-photometry (red dots), and the result of our 
  spectral synthesis with a two-temperature model (see text for details)
  Whenever possible, orbital minimum and maximum brightness are
  indicated and connected by lines}
\label{f:seduvopt}
\end{figure}

The UVW1-filter is centered on 2910\AA, between the GALEX-NUV passband
(Szkody et al.~2006) and the optical broad-band filters. 
Mean flux values as measured with the OM are shown together with other
low-state photometric and spectroscopic data (Harrison et al.~2004, Szkody et
al.~2006) from the infrared to the ultraviolet spectral range in
Fig.~\ref{f:seduvopt}. It shows that the different low state observations
are compatible with each other. 

Based on an analysis of their low-state {\it GALEX} ultraviolet and optical 
photometry Szkody et al.~(2006) state the existence of a light-source
reminiscent of a 20000\,K hot spot. Their spot model, however, could neither 
explain the large-amplitude FUV  variations nor the spectral energy
distribution 
and they arrive at the conclusion, that {\it no} spot model can
explain their observations.

The analysis by Szkody et al.~(2006) was based on an assumed effective
temperature $\teff = 9500$\,K (Beuermann et al.~2000) which they approximated
as a Planckian function. We note, that this approximation indeed gives
rise to a large ultraviolet excess. We re-address the question of the white
dwarf and spot temperature making use of state-of-the-art white
dwarf model atmospheres (Koester et al.~2005 and references therein).
The optical spectrum alone is best described by a model with $\teff =
10500\pm1000$\,K. This value is in accord with the more recent analysis by the
G\"ottingen group (Beuermann et al.~2007) who used $\teff = 11000 \pm 1500$\,K.

The pure white dwarf model falls short of matching the observed ultraviolet
flux. We therefore fitted the SED at orbital minimum and maximum 
and the ultraviolet/optical light curves 
with a two-temperature model, a cooler one representing the white dwarf and a
hotter one representing a spot. We folded our white-dwarf model spectra
with the effective area curves of the two {\it GALEX}
passbands\footnote{http://galexgi.gsfc.nasa.gov/tools/Resolution\_Response/index.html}
thus converting Eddington flux to count rate. 
Size, temperature and location of the spot and the temperature of the white
dwarf were varied until a satisfactory fit (by eye) to the optical and ultraviolet light
curves and the SED were reached. We 
arrived at a consistent solution for $T_{\rm wd} = 9750$\,K and $T_{\rm spot}
= 18500$\,K with an estimated uncertainty of 1000\,K in the white dwarf
temperature. The spot temperature is subject to much larger systematic
uncertainties, since e.g.~our assumption of a uniform spot temperature is a
very crude approximation. The model spectra shown in Fig.~\ref{f:seduvopt}
at orbital minimum and maximum were computed for a binary inclication
$i=60\degr$, a spot extent of $24\degr$ (half opening angle), and a spot
colatitude of just $12.5\degr$. 
A rather high inclination and a high 'northern' spot latitude,
the spot undergoes a partial self-eclipse only, are required from the fact that
the FUV-band is almost completely dominated by the spot. Even at orbital
minimum the white dwarf contributes only $\sim$10\% to the total flux in that
band (see the lowest model curve in Fig.~\ref{f:seduvopt}).
Our model is simple and far from being unique but 
it fits the data well and contradicts the 
conclusion by Szkody et al.~(2006)
that no spot model can explain both the SED and the variability. 

Combining parallaxes, proper motions and absolute magnitude constraints,
Thorstensen (2003) derived distance estimates for 14 CVs with a Bayesian
method, among them \ef. Dependent on the proper-motion, magnitude and velocity
priors, he derived a short, $d = 113^{+19}_{-16}$\,pc, and a long, $d =
163^{+66}_{-50}$\,pc, distance to \ef. At 113\,pc the observed flux of our
9750 K white dwarf model results in a radius of $6.5 \times 10^{8}$\,cm and
implies a relatively massive white dwarf of 0.87\,\msun. At 163\,pc the implied
mass is $\sim$0.55\,\msun. These numbers differ slightly from those derived in 
Beuermann et al.~(2000), since the spot contribution was taken into account
in our analysis.

\section{Results and discussion}
We have analysed archival XMM-Newton observations of \ef\ obtained in 2002 and
2003. At all three occasions the polar was detected as an X-ray source,
although at a very low flux level. The spectra were compatible with emission 
from a low-temperature corona-like plasma. The longest X-ray observation had
almost full phase coverage, the plasma temperature was as low as 2\,keV or
less. The mean orbital integrated flux in this component is about $F_X =
6 \times 10^{-14}$\ftot. Assuming isotropic radiation and a distance of 
163\,pc (Thorstensen 2003), a luminosity of 
$L_X \sim 2 \times 10^{29}$\,\lum\  is derived. 

The question arises if this faint X-ray flux originates from the corona of the
secondary or from the accretion region on the white dwarf. Neither X-ray
variability nor the X-ray spectrum give a clear answer. Both, 
a coronal plasma and the cooling plasma from low level accretion have
such low temperatures as measured here. Evidence for X-ray emission from an
accretion plasma can be given indirectly. Firstly, although not very much is
known about X-ray emission from degenerate stars at the bottom of the main
sequence, their X-ray luminosities seem to fall short by one dex with respect
to the X-ray luminosiy of \ef\,(Stelzer et al.~2006). Secondly, \ef\ shows
clear signs of residual accretion via the detection of infrared cyclotron
harmonics (Harrison et al.~2004). It therefore appears reasonable to
  assign the observed X-ray emission to some remaining weak accretion. This
  will be our working hypothesis in the following.
It remains unclear if residual accretion happens via an accretion
stream or via a 
stellar wind, the latter being inferred in order to explain the faint X-ray
emission from the small group of pre-CVs (termed also LARPs, Schwope et
al.~2002, Schmidt et al.~2005, Vogel et al.~2006). Since the emission region
in \ef\ is not self-eclipsing (Beuermann et al.~1987, 1991), we are lacking a
distinct photometric feature to discern between the two possibilities. 
The pronounced soft X-ray absorption dip 
as a sign of stream accretion and
seen in high accretion states was not 
secularly detected here, but due to the small number of photons its
absence does not give a clear-cut answer to the question of the accretion
mode.

\begin{figure}[t]
\resizebox{\hsize}{!}{\includegraphics[clip=]{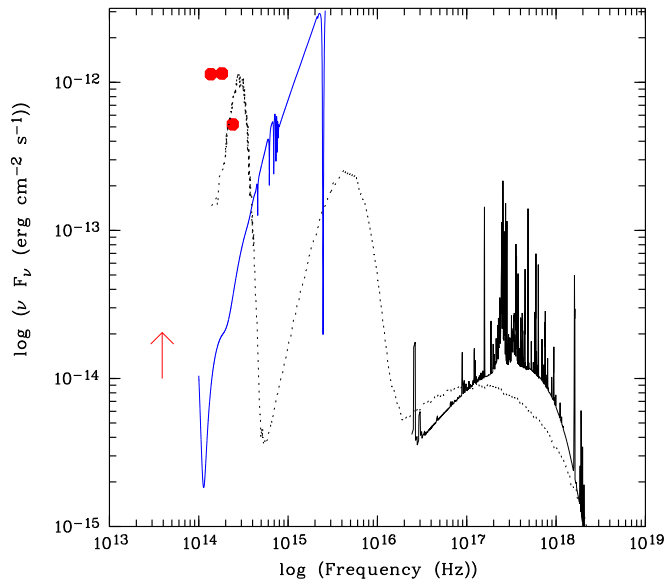}}
\caption{Infrared to X-ray spectral energy distribution of \ef\ in  the
  low state. Shown are radiation components which are associated with the
  accretion process, i.e.~corrected for stellar photospheric radiation, at
  orbital maximum. The  red arrow indicates the cyclotron fundamental for
  $B=14$\,MG.} 
\label{f:sedirx}
\end{figure}

We discuss the energy balance of the accretion process in the low state
on the assumption that the observed X-ray emission is due to accretion onto
the white dwarf primary. We make the further assumption that the excess
emission in the infrared over the extrapolated white dwarf spectrum is solely
due to cyclotron emission from the accretion plasma.
The relevant radiation components are shown in
Fig.~\ref{f:sedirx}. 
It shows the coronal plasma in the
X-ray regime and the cyclotron component in the infrared, the latter corrected
for the contribution from the underlying white dwarf. 
Included in the figure is the spot model at orbital maximum, represented by a
white dwarf model spectrum with $\teff = 18500$\,K.

The integrated flux in the thermal plasma X-ray component is $F_X = 6\times
10^{-14}$\,\ftot. At an assumed field strength of 13\,MG (Beuermann et
al.~2007) the cyclotron fundamental is at about $8\mu$m. If we assume a 
Rayleigh-Jeans limited flux up to the $H$-band, where the cyclotron component
peaks (Harrison et al.~2004 and Figs.~\ref{f:seduvopt} and \ref{f:sedirx}),
the integrated cyclotron flux is 
about $F_{\rm cyc} \simeq 4\times 10^{-12}$\,\ftot. 
This value has an uncertainty of at least 50\%, since the
cyclotron spectrum is covered only partly by observations. 

There is no evidence for any component of re-processed radiation in the
extreme ultraviolet/soft X-ray regime. We assume that the 
spot component in the ultraviolet carries the complete
information on reprocessed radiation. The integrated flux in this component is
of order $F_{\rm rep} \simeq 2 \times 10^{-11}$\,\ftot. 
This is clearly larger than the primary cyclotron radiation and a
factor $\sim$300 larger than the thermal X-ray component. The large amount of
radiation in the ultraviolet in excess of the primary radiation components
is suggestive of a reservoir of heat from the previous high state and does not
support a picture of instantaneous reprocessing. Clearly, more UV-observations
are necessary to verify this picture by determining the cooling curve of
the accretion spot.

We thus derive a flux balance $F_{\rm UV} \simeq 5 \times F_{\rm cyc}$ and 
$F_{\rm cyc} \sim 60 F_X$, the latter ratio being 
suggestive of accretion in the bombardement regime at low mass flow rates
and/or high magnetic field. The field of \ef\ is one of the lowest among all
polars which made the binary a rather hard X-ray source with almost balanced
flux contributions in the high state (see the detailed discussion in BSP87).
Since Beuermann et al. found \ef\ to
behave as described in the 'standard' accretion scenario (Lamb \& Masters
1979), it was termed 'the textbook example of AM Herculis stars' by them. 
This picture changed fundamentally in the low state. 

Beuermann (2004) presented a sequence of model spectra for $B=14$\,MG, $\Theta
= 60\degr$ and variable mass flow rate $\dot{m}$ (in
g\,cm$^{-2}$\,s$^{-1}$). The values of $B$ and $\Theta$ are quite similar to
those of \ef. We include his model for $\dot{m} 
=10^{-2}$\,g\,cm$^{-2}$\,s$^{-1}$ in Fig.~\ref{f:sedirx}. This model predicts
the right (within an order of magnitude) flux ratio between cyclotron and
X-ray flux. At even lower mass flow rates, the next smaller value computed by
Beuermann (2004) is $\dot{m}
=10^{-3}$\,g\,cm$^{-2}$\,s$^{-1}$, the flux ratio $F_{\rm cyc}/F_X$
becomes much smaller than observed. Also, the predicted size of the cyclotron
emitting area would be larger than the white dwarf (for an assumed distance of
120\,pc). We thus regard $\dot{m} = 10^{-2}$\,g\,cm$^{-2}$\,s$^{-1}$ as the
likely value for \ef\ in its low accretion state. Fig.~\ref{f:sedirx} shows
the blackbody approximation as non-appropriate for the reprocessed
component. This was noted already by Beuermann (2004), the low state
observations of \ef\ prove this observationally. 

At $\dot{m} = 10^{-2}$\,g\,cm$^{-2}$\,s$^{-1}$ and $B=13$\,MG the maximum
predicted electron temperate is about 7\,keV (Fischer \& Beuermann 2001). 
Our measured temperature $kT \simeq 1.7$\,keV (E501) indicates that the bulk
of X-ray emission originates from denser layers at lower temperatures, as
expected.

A comparison between high and low states fluxes of the main radiation
components is instructive. For the mean fluxes in the high state, BSP87 
derive $F_{\rm cyc} = 4.8 \times 10^{-11}$\,\ftot,  
$F_{\rm brems} = 1.5 \times 10^{-10}$\,\ftot, and
$F_{\rm bb} = 5.5 \times 10^{-10}$\,\ftot, respectively. 
Hence, when switching from the high to the low state, the cyclotron flux is
reduced by a factor $\sim$10, 
and the flux in the thermal plasma component by a factor
2500. These numbers illustrate the variable occupation of the different
channels of energy release, when switching from a high-accretion rate,
shock-dominated flow to the low-rate, cyclotron-dominated bombarded
atmosphere. A direct comparison between the high- and
low-state fluxes in the re-processed component seems not to be possible, since
a counterpart to the high-state blackbody is missing in the low
state. The low-state ultraviolet flux seems still to be fed high-state
accretion heating.

The bolometric flux in the low state is 
$F_{\rm acc} \simeq F_{\rm cyc} \simeq 4 \times 10^{-12}$\,\ftot, 
the luminosity is $L_{\rm acc} \simeq 2 \pi d^2 F_{\rm acc}
\simeq 2.4 \times 10^{30} d_{100}^2$\,\lum, a factor 100 -- 300 lower than the
high state accretion luminosity derived by BSP87. 

We may also compare the cyclotron emitting areas in the high and low
states. BSP87 derive $A_{\rm cyc} = (3.2-12.2) \times 10^{15}
d_{100}^2$\,cm$^{2}$ while scaling of the model shown in Fig.~\ref{f:sedirx}
implies $A_{\rm cyc} \simeq 30 \times 10^{15} d_{100}^2$\,cm$^{2}$. Again,
care has to be taken by taking the latter number too literally since we have
just scaled a pre-existing model, but they imply that the cyclotron emitting
area has not shrunk by orders of magnitude. 

A final comment may be made on \ef\ as a polar and its relation to the class
of (likely misleadingly termed) LARPs (Low-Accretion Rate Polars; Schwope et 
al.~2002). The latter are close white 
dwarf/M dwarf pairs with pronounced cyclotron harmonics which led to their
discovery in the HQS and the SDSS spectroscopic surveys. They are likely
detached binaries acrreting from the stellar wind of the secondary (Schmidt et
al.~2005, Vogel et al.~2007). \ef\ is in a deep low state now for about 10
years. It 
is faint, its optical spectrum shows just the white dwarf and no secondary,
variability in the optical is small, $\Delta V < 0.3^m$, its X-ray flux is
faint and below all flux limits of surveys with sufficiently large survey
area. Hence, all the classical discovery channels of CVs would not have led
to the identification of \ef\ as a rather nearby cataclysmic variable in its
extended low state. 
It would have been classified just as an isolated 
magnetic white dwarf in a spectroscopic survey. 

But it shows an infrared excess, which could be (erroneously) 
interpreted as originating from a secondary and therefore would hint to the
binary nature of this source. 
Actually it still shows a high degree of variability, 
although most pronounced in the ultraviolet. 
And finally, it still emits X-rays, but at the given flux the X-ray sky is
dominated by AGNs. Hence, it still shows many hallmarks of the polars.
and in that respect we may 
coin \ef\ as the textbook example of the low-accretion rate polars.
We are not going to speculate about a population of
missing CVs in similar extended low states. 
But the case of
\ef\ as a secure low-accretion rate polar underlines the importance of a
multi-wavelength approach to find more of these intriguing sources.

\begin{acknowledgements}
We thank our referee, Klaus Beuermann, for helpful comments which improved the
manuscript. 
We thank V. Hambaryan and G. Lamer for help with the data reduction.
This work was supported in part by the Deutsches Zentrum f\"ur Luft- und
Raumfahrt (DLR) GmbH under contract No. FKZ 50 OR 0404 and by the Deutsche
Forschungsgemeinschaft DFG under contract No.~Schw536/20-1.
\end{acknowledgements}

\end{document}